\documentclass{amsart} \usepackage{amsmath,amssymb,latexsym,amscd}   \newcommand{\e}{\epsilon}  \newcommand{\va}{\vartheta} \newcommand{\tva}{\tilde{\vartheta}}   
\newcommand{\beq }{\begin{equation}} \newcommand{\eeq }{\end{equation}} \newtheorem{theorem}{Theorem}  
  \theoremstyle{definition} \newtheorem{remark}{Remark} 
\numberwithin{equation}{section}

\begin{document} \title[Supplementary balance laws] {\textbf{Supplementary balance laws for Cattaneo heat propagation.}} \author{Serge Preston}(\address{Department
of Mathematics and Statistics, Portland State University, Portland, OR, U.S.}) \email{serge@mth.pdx.edu}. \vskip0.4cm \date{} \maketitle

\begin{abstract} In this work we determine for the Cattaneo heat propagation system  all the supplementary balance laws (conservation laws )  of the same order
(zero) as the system itself and extract the constitutive relations (expression for the internal energy) dictated by the Entropy Principle. The space of all
supplementary balance laws (having the functional dimension 8) contains four original balance laws and their deformations depending on 4 functions of temperature
($\lambda^0(\va), K^A(\va),\ A=1,2,3 $).  The requirements of the II law of thermodynamics leads to the exclusion of three functional degrees ($K^A=0,\ A=1,2,3$) and
to further restriction to the form of internal energy. In its final formulation, entropy balance represent the deformation of the energy balance law by the
functional parameter $\lambda^{0}(\va)$. \end{abstract}

\today \section{Introduction.} In this work we determine the form of all supplementary balance laws for the Cattaneo heat propagation system (CHP-system) (2.1)
bellow. We will solve the LL-equations \cite{MR,Pr1} directly, and get the constitutive relation on the internal energy as the function of temperature $\theta$ and
heat flux $q$. If this condition is fulfilled, the total space of SBL (modulo trivial balance laws) is 8-dimensional, if this condition does not hold, there are no
new SBL. Then we show that the positivity condition for the production in the new balance laws place additional restriction to the form of internal energy and
determine the unique SBL having nonnegative production - entropy balance law. \section{Supplementary balance laws of a balance system.}

Let one has a system of balance equations for the fields $y^i (t,x^A, A=1,2,3)$ \beq \partial_{t}F^{0}_{i}+\partial_{X^A}F^{A}_{i}=\Pi_i, \ i=1,\ldots,m \eeq with
the densities $F^{0}_{i}$, fluxes $F^{A}_{i},A=1,2,3$ and sources $\Pi_{i}$  being functions of space-time point $(t,x^A, A=1,2,3)$, fields $y^i$ and their
derivatives (by $t,x^A$) up to the order $k\geqq 0$.  Number $k$ is called the order of balance system (1.1).  In continuum thermodynamics people mostly work with
the balance system of order $0$ (case of Rational Extended Thermodynamics) and $1$.  \par

A balance law of order $r$ (in the same sense as the system (1.1) is of order $k$). \beq \partial_{t}K^{0}+\sum_{A=1}^{3}\partial_{x^A}K^{A}=Q \eeq is called a
\textbf{supplementary balance law for the system (SBL) (1.1)} if \textbf{any solution of the system satisfies to the balance equation (1.2)}. \par Examples of
supplementary balance laws are: entropy balance, provided the Entropy Principle is admitted for system (1.1), see \cite{R1,MR,R3}, Noether symmetries in the sense of
works \cite{Pr,Pr1} and some linear combinations of the balance equations of system (1.1) satisfying to some condition (gauge symmetries of system (1.1), see
\cite{Pr, Pr1} ).\par As a rule, in classical physics one looks for entropy balance laws of the same order as the original balance systems.  Higher order SBL are
also of an interest for studying the system (1.1) -  for example, study of integrable systems leads to the hierarchy of conservation laws (often having form of
conservation laws themselves) of higher order. \par

For a balance systems (1.10 of order $0$ (case of Rational extended Thermodynamics, \cite{MR}), density and flux of a SBL (1.2) satisfy to the system of equations
\beq \lambda^i F^{\mu}_{i,y^j}=K^{\mu}_{,y^j},\ \eeq where summation by repeated indices is taken. Functions $\lambda^{i}(y^j)$ (main fields in terminology of
\cite{MR}) are to be found from the conditions of solvability of this system.  We call this system - the LL-system refereing to the Liu method of using Lagrange
method for formulating dissipative inequality for a system (1.1), see \cite{L,MR}.  Source/production in the system (1.2) is then found as $Q=\sum_{i}\lambda^i
\Pi_{i}$.

\section{Cattaneo Heat propagation balance system.}

\par Consider the heat propagation model containing the temperature $\vartheta$ and heat flux $q$ as the independent dynamical fields. $y^0 =\vartheta, y^A=q^A,\
A=1,2,3$. \par

Balance equations of this model have the form

\beq \begin{cases} \partial_{t}(\rho \epsilon)+div(q)=0,\\ \partial_{t}(\tau q)+ \nabla \Lambda(\vartheta) =-q. \end{cases} \eeq Second equation can be rewritten in
the conventional form \[\partial_{t}(\tau q)+ \lambda \cdot\nabla \vartheta =-q,\] where $\lambda =\frac{\partial \Lambda }{\partial \vartheta}.$  If coefficient
$\lambda$ may depend on the density $\rho$, equation is more complex. \par Constitutive relation specify dependence of the internal energy $\epsilon$ on $\vartheta
,q$ and possible dependence of coefficients $\tau, \Lambda$ on the temperature (including the requirement $\Lambda_{,\vartheta}\ne 0$).  Simplest case is the linear
relation $\epsilon =k\vartheta$, but for our purposes it is too restrictive, see \cite{JCL}, Sec.2.1.\par Since $\rho$ is not considered here as a dynamical
variable, we merge it with the field $\epsilon$ and from now on and till the end it will be omitted. On the other hand, in this model the the energy $\epsilon$
depends on temperature $\vartheta$ \emph{and on the heat flux} $q$ (see \cite{JCL},Sec.2.1.2) or, by change of variables, temperature $\vartheta=\vartheta(\epsilon,
q)$ will be considered as the function of dynamical variables.\par

Cattaneo equation $q+\tau \partial_{t}(q)=-\lambda\cdot \nabla \vartheta$ has the form of the vectorial balance law and, as a result there is no need for the
constitutive relations to depend on the derivatives of the basic fields.  No derivatives appears in the constitutive relation, therefore, this is the RET model. In
the second equation there is a nonzero production $\Pi^A=-q^A$. Model is homogeneous, there is no explicit dependence of any functions on $t,x^A$.\par

\section{LL-system for supplementary balance laws of CHP-system} To study the LL-system for the supplementary balance laws we start with the $i\times \mu$ matrix of
density/flux components

\[ F^{\mu}_{i}=\begin{pmatrix} \epsilon & \tau q^1 & \tau q^2 & \tau q^3\\
 q^1 & \Lambda(\vartheta) & 0 & 0\\
 q^2 &0 & \Lambda(\vartheta) & 0\\
 q^3 & 0 & 0 & \Lambda(\vartheta)
\end{pmatrix}. \] Assuming that coefficients $\tau$ and the function $\Lambda$ are independent on the heat flux variables $ q^A$ we get the "vertical (i. e by fields
$\vartheta ,q^A$) differentials" of densities and flux components $F^{\mu}_{i}$ \[ d_v F^{\mu}_{i}=\begin{pmatrix} \e_{\va}d\va +\e_{q^A}dq^A  & \tau_{\va}q^1 d\va+
\tau dq^1 &\tau_{\va}q^2 d\va+\tau dq^2 &\tau_{\va}q^2 d\va+\tau dq^3\\
 dq^1 & \Lambda_{\va} d\vartheta & 0 & 0\\
 dq^2 &0 & \Lambda_{\va} d\vartheta & 0\\
 dq^3 & 0 & 0 &\Lambda_{\va} d\vartheta
\end{pmatrix}. \] \par Let now \beq \partial_{t}K^{0}(x,y)+\partial_{x^A}K^{A}(x,y)=Q(x,y) \eeq be a supplementary balance law for the Cattaneo balance system (2.1).
It is easy to see that the LL-system has the form As a result, LL subsystem for $\mu =0$ takes the form \vskip0.4cm

For $\mu=0$, \beq \begin{cases} \lambda^0 \e_{\va}+\tau_{\va}\lambda^A q^A  =K^0_{,\va},\\ \ \lambda^0 \e_{q^A}+\lambda^A \tau  =K^{0}_{,q^A}\end{cases},\ A=1,2,3 .
\eeq \vskip0.4cm For $A=1,2,3$, using cyclic notations, we have LL-equations

\beq \begin{cases} \Lambda_{\va} \lambda^A  =K^A_{,\va},\\ \ \lambda^0  =K^{A}_{,q^A},\\ 0=K^{A}_{q^{A+1}},\\ 0=K^{A}_{q^{A+2}},\ A=1,2,3 .\end{cases} \eeq

Looking at systems (4.2-4.3) we see that if we make the change of variables: $\tva =\Lambda (\va )$ then the system of equations (4.2-3)takes the form ( wherever is
the derivative by $\va$ we multiply this equation by $\Lambda_{,\va}$)

\beq \begin{cases} \lambda^0 \e_{\tva}+\tau_{\tva}\lambda^A q^A  =K^0_{,\tva},\\ \ \lambda^0 \e_{q^A}+\lambda^A \tau  =K^{0}_{q^A}\end{cases};\ \begin{cases}
K^A_{,\tva}=\lambda^A,\\ \ K^{A}_{q^B}=\lambda^0 \delta^{A}_{B}\end{cases},\ A,B=1,2,3 . \eeq Second subsystem is equivalent to the relation \[ d_{v}K^A=\lambda^A
d\tilde{\vartheta} +\lambda^0 dq^A. \]

These integrability conditions imply the expression $K^A =K^A (x^\mu, \tva ,q^A)$ and \[ K^{A}_{q^A}=\lambda^{0},\ A=1,2,3\ \Rightarrow \lambda^0 =\lambda^0 (\tva).
\] Integrating equation $K^{A}_{q^A}=\lambda^{0}(\tva)$ by $q^Q$ we get \beq K^{A}=\lambda^{0}(\tva)q^A +\tilde{K}^{A}(\tva) \eeq with some functions
$\tilde{K}^{A}(\tva).$\par First equation of each system now takes the form \beq \lambda^A =K^{A}_{\tva}=\lambda^{0}_{\tva}q^A +\tilde{K}^{A}_{,\tva}(\tva). \eeq
Substituting these expressions for $\lambda^A$ into the 0-th system \[
 \begin{cases} \lambda^0 \e_{\tva}+\tau_{\tva}\lambda^A q^A  =K^0_{,\tva},\\ \ \lambda^0 \e_{q^A}+\lambda^A \tau  =K^{0}_{q^A}\end{cases},\ A=1,2,3,
\] we get \beq
 \begin{cases} K^0_{,\tva}=\lambda^0 \e_{\tva}+\tau_{\tva}(\lambda^{0}_{\tva}\Vert q\Vert^2 +\tilde{K}^{A}_{,\tva}(\tva)q^A)  ,\\ \ K^{0}_{q^A}=\lambda^0
 \e_{q^A}+\tau
 (\lambda^{0}_{\tva}q^A +\tilde{K}^{A}_{,\tva}(\tva)) \end{cases},\ A=1,2,3,
\eeq where $\Vert q\Vert^2 =\sum_{A}q^{A\ 2}$.\par

Integrating $A$-th equation by $q^A$ and comparing results for different $A$ we obtain the following representation \beq K^0 =\lambda^0 \e +\tau(\tva
)[\frac{1}{2}\lambda^{0}_{\tva}\Vert q\Vert^2+\tilde{K}^{A}_{,\tva}(\tva)q^A]+f(\tva) \eeq for some function $f(\tva,x^\mu )$.\par

Calculate derivative by $\tva$ in the last formula for $K^0$ and subtract the first formula of the previous system.  We get \beq 0=\lambda^{0}_{,\tva} \e +\tau(\tva
)[(\frac{1}{2}\lambda^{0}_{\tva}\Vert q\Vert^2+\tilde{K}^{A}_{,\tva}(\tva)q^A)]_{,\tva}-\frac{1}{2}\tau_{,\tva}\lambda^{0}_{,\tva}\Vert q\Vert^2  +f_{,\tva}(\tva).
\eeq This is the compatibility condition for the system (4.2) for $K^0$.  As such, it is realization of the general compatibility system (4.4).\par

Take $q^A=0$ in the last equation, i.e. consider the case where \emph{there are no heat flux}.  Then the internal energy reduces to its \emph{equilibrium value }
$\e^{eq}(\tva)$ and we get $f_{,\tva}(\tva)=-\lambda^{0}_{,\tva} \e^{eq}$.  Integrating here we find \beq f(\tva )=f_{0}(x^\mu)-\int^{\tva}\lambda^{0}_{,\tva}(s)
\e^{eq}(s) ds. \eeq Substituting this value for $f$ into the previous formula and we get expressions for $K^\mu$:

\beq \begin{cases} K^0 =\lambda^0 \e -\int^{\tva}\lambda^{0}_{,\tva} \e^{eq} ds +\tau(\tva )[\frac{1}{2}\lambda^{0}_{\tva}\Vert
q\Vert^2+\tilde{K}^{A}_{,\tva}(\tva)q^A]+f_{0},\\ K^{A}=\lambda^{0}(\tva)q^A +\tilde{K}^{A}(\tva),\ A=1,2,3. \end{cases} \eeq In addition to this, from (4.9) and
obtained expression for $f(\tva)$, we get the \emph{expression for internal energy} \beq \e=\e^{eq}(\tva)+\frac{1}{2}\tau_{,\tva }\Vert q\Vert^2-\frac{\tau
(\tva)}{\lambda^{0}_{\tva}(\tva)}\left[ \frac{1}{2}\lambda^{0}_{,\tva \tva}\Vert q\Vert^2+\tilde{K}^{A}_{,\tva\tva}(\tva)q^A\right]. \eeq \textbf{This form for
internal energy present the restriction to the constitutive relations in Cattaneo model placed on it by the entropy principle.}\par

Zero-th main field $\lambda^0$ is an arbitrary function of $\tva $ while $\lambda^A$ are given by the relations (4.11): \beq \lambda^A = (\lambda^{0}_{\tva}q^A
+\tilde{K}^{A}_{,\tva}(\tva)). \eeq Using this we find the source/production term for the SBL (4.1) \beq Q=\lambda^A \Pi_{A}=-\lambda^A q^A
=-(\lambda^{0}_{\tva}\Vert q\Vert^2 +\tilde{K}^{A}_{,\tva}(\tva)q^A). \eeq

Now we combine obtained expressions for components of a secondary balance law.  We have to take into account that the LL-system defines $K^\mu $ only $  mod \
C^{\infty}(X)$.  This means first of all that all the functions may depend explicitly on $x^\mu$. For energy $\e$, field $\Lambda(\va )$ and the coefficient $\tau$
this dependence is determined by constitutive relations and is, therefore, fixed. Looking at (4.12) we see that the coefficients of terms linear and quadratic by
$q^A$ are also defined by the constitutive relation, i.e. in the representation \beq \e=\e^{eq}(\tva)+\mu(\tva )\Vert q\Vert^2+ M_{A}(\tva )q^A =
\e^{eq}(\tva)+\frac{1}{2}\tau_{,\tva }\Vert q\Vert^2-\frac{\tau (\tva)}{\lambda^{0}_{\tva}(\tva)}\left[ \frac{1}{2}\lambda^{0}_{,\tva \tva}\Vert
q\Vert^2+\tilde{K}^{A}_{,\tva\tva}(\tva)q^A\right], \eeq coefficients \beq \mu(\tva ,x)=\frac{1}{2}\tau_{,\tva }-\frac{1}{2}\frac{\tau
(\tva)}{\lambda^{0}_{\tva}(\tva)}\lambda^{0}_{\tva\tva},\ M_{A}=-\frac{\tau (\tva)}{\lambda^{0}_{\tva}(\tva)}\tilde{K}^{A}_{,\tva\tva}(\tva) \eeq are defined by the
CR - by expression of internal energy as the quadratic function of the heat flux.\par More then this, quantities $\frac{\lambda^{0}_{\tva\tva}}{\lambda^{0}_{\tva}}$
and $\frac{\tilde{K}^{A}_{,\tva\tva}(\tva)}{\lambda^{0}_{\tva}}$ are also defined by the constitutive relations.\par Rewriting the first relation (14.15) we get
\begin{multline} \left(ln(\lambda^{0}_{\tva})\right)_{,\tva}=ln(\tau )_{,\tva} -2\frac{\mu(\tva )}{\tau (\tva)}  \Rightarrow
ln(\lambda^{0}_{\tva})=ln(\tau)+b^0-2\int^{\tva}\frac{\mu}{\tau}(s)ds \Rightarrow \\ \Rightarrow \lambda^{0}_{\tva}=\alpha \tau
e^{-2\int^{\tva}\frac{\mu}{\tau}(s)ds},\ \alpha=e^{b^0}>0. \end{multline} From this relation we find \beq
 \lambda^{0}(\tva,x)=a^0+\alpha\hat{\lambda}^{0}=a^0+\alpha \int^{\tva}[\tau e^{-2\int^{u}\frac{\mu (s)}{\tau (s)}ds}]du
\eeq Here $a^0$ and $\alpha$ are constants (or, maybe, functions of $x^\mu$(?).\par Using obtained expression for $\lambda^{0}(\tva,x)$ in the second formula (3.16)
we get the expression for coefficients $\tilde{K}^{A}$ and, integrating twice by $\tva$, for the functions $K^{A}(\tva)$ \begin{multline}
\tilde{K}^{A}_{,\tva\tva}=-M_{A}\cdot \frac{\lambda^{0}_{\tva}(\tva)}{\tau (\tva)}=-M_{A}\alpha e^{-2\int^{\tva}\frac{\mu}{\tau}(s)ds}\Rightarrow \\ \Rightarrow
\tilde{K}^{A} =k^A \tva +m^A +\alpha \cdot \hat{K}^{A}(\tva ) =k^A \tva +m^A-\alpha \int^{\tva} dw\int^{w}[M_{A}(u)e^{-2\int^{u}\frac{\mu}{\tau}(s)ds}]du.
\end{multline}

Functions $\hat{K}^{A}(\tva )$ are defined by the second formula in the second line.\par Thus, functions $\lambda^{0}_{\va},\tilde{K}^{A}_{,\va\va}$ are defined by
the constitutive relations while coefficients $\alpha >0,a^0,k^A,m^A$ are arbitrary functions of $x^\mu$. \par \section{Supplementary balance laws for CHP-system.}
Combine obtained results, returning to the variable $\va$  (and using repeatedly the relation $f_{,\tva}=\va_{,\tva} f_{,\va}=(\tva_{,\va})^{-1}
f_{,\va}=\Lambda^{-1}_{,\va}f_{,\va}$) we get the general expressions for admissible densities/fluxes of the supplementary balance laws \beq \begin{cases} K^0
=\lambda^0 \e -\int^{\tva}\lambda^{0}_{,\tva} \e^{eq} ds +\tau(\tva )[\frac{1}{2}\lambda^{0}_{\tva}\Vert
q\Vert^2+\tilde{K}^{A}_{,\tva}(\tva)q^A]+f_{0}=\\=(a^0+\alpha\hat{\lambda}^0 ) \e -\alpha \int^{\va}\hat{\lambda}^{0}_{,\va} \e^{eq} ds +\frac{\tau(\va
)}{\Lambda_{,\va}}[\frac{\alpha}{2}\hat{\lambda}^{0}_{\va}\Vert q\Vert^2+(\Lambda_{,\va} k^A+\alpha\hat{K}^{A}_{,\va}(\va))q^A]+f_{0},\\ K^{A}=\lambda^{0}(\tva)q^A
+\tilde{K}^{A}(\tva)=(a^0+\alpha\hat{\lambda}^{0}(\va))q^A + k^A \Lambda(\va) +m^A +\alpha\hat{K}^{A}(\va ) ,\ A=1,2,3.\\ Q=-\lambda^A q^A =-(\lambda^{0}_{\tva}\Vert
q\Vert^2 +\tilde{K}^{A}_{,\tva}(\tva)q^A)=-\Lambda^{-1}_{,\va}(\lambda^{0}_{\va}\Vert q\Vert^2 +\Lambda_{,\va } k^A q^A +\alpha\hat{K}^{A}_{,\va}(\va)q^A )=\\
-\Lambda^{-1}_{,\va}(\alpha \hat{\lambda}^{0}_{\va}\Vert q\Vert^2 +\Lambda_{,\va } k^A q^A +\alpha\hat{K}^{A}_{,\va}(\va)q^A ). \end{cases} \eeq

Collecting previous results together we present obtained expressions for secondary balance laws first in short form and then - in the form where original balance
laws and the trivial balance laws are separated from the general form of SBL \begin{multline} \begin{pmatrix}K^0\\K^1\\ K^2\\ K^3 \\ Q\end{pmatrix} =
\begin{pmatrix}\lambda^0 \e -\int^{\va}\lambda^{0}_{,\va} \e^{eq} ds +\tau(\va )\Lambda^{-1}_{\va}[\frac{1}{2}\lambda^{0}_{\va}\Vert q\Vert^2+\alpha
\tilde{K}^{A}_{,\va}(\va)q^A]+f_{0}\\ \lambda^{0}(\va)q^1 +\tilde{K}^{1}(\va)\\ \lambda^{0}(\va)q^2 +\tilde{K}^{2}(\va)\\ \lambda^{0}(\va)q^3 +\tilde{K}^{3}(\va)\\
-\Lambda^{-1}_{,\va }(\lambda^{0}_{,\va}\Vert q\Vert^2 +\tilde{K}^{A}_{,\va}(\va)q^A)\end{pmatrix}=\\=a^0 \begin{pmatrix} \e \\ q^1\\q^2\\q^3 \\0 \end{pmatrix}+
\sum_{A}k^A \begin{pmatrix} \tau(\va )q^A \\ \delta^{1}_{A} \Lambda( \va) \\ \delta^{2}_{A} \Lambda(\va) \\ \delta^{3}_{A} \Lambda(\va) \\ - q^A\end{pmatrix}
+\begin{pmatrix} \alpha \tau \Lambda(\vartheta)^{-1}\hat{K}^{A}_{,\va}(\va)q^A\\ \hat{K}^{1}(\va)\\ \hat{K}^{2}(\va)\\ +\hat{K}^{3}(\va) \\ -\Lambda^{-1}_{,\va
}\hat{K}^{A}_{,\va}(\va)q^A \end{pmatrix} + \alpha \begin{pmatrix}\hat{\lambda}^0 \e -\int^{\va}\hat{\lambda}^{0}_{,\va} \e^{eq} ds +\tau(\va
)\Lambda^{-1}_{\va}[\frac{1}{2}\hat{\lambda}^{0}_{,\va}\Vert q\Vert^2]\\ \hat{\lambda}^{0}(\va)q^1 \\ \hat{\lambda}^{0}(\va)q^2 \\ \hat{\lambda}^{0}(\va)q^3 \\
-\Lambda^{-1}_{,\va }\hat{\lambda}^{0}_{,\va}\Vert q\Vert^2 \end{pmatrix}+\begin{pmatrix} f_0\\ m^1\\m^2\\m^3 \\0 \end{pmatrix}. \end{multline}

To get the second presentation of the SBL we use the decompositions (4.19) $\lambda^0=\alpha\hat{\lambda}_{0}+a^0$ and (4.18) $\tilde{K}^{A}(\tva)=k^A \tilde
\vartheta +m^A-\hat K^A.$

\begin{remark} Notice the duality between the tensor structure of the basic fields of Cattaneo system - one scalar field (temperature $\va$) and one vector field
(heat flux $q^A, A=1,2,3$) and the structure of space $\mathcal{SBL}(C)$ of supplementary balance laws - elements of $\mathcal{SBL}(C)$ depend on one scalar function
of temperature $\lambda^{0}(\va)$ and one covector function of temperature $\hat{K}_{A}$. \end{remark}

\begin{remark} It is easy to see that none of new SBL can be written as a linear combination of original balance equations with variable coefficients (Noether
balance laws generated by vertical symmetries $v=v^k(y^i)\partial_{y^k}$, see \cite{Pr,Pr1} ). Easiest way to prove this is to compare the source terms of different
balance equations. \end{remark} Returning to the variable $\va$ in the expression (3.12) and using the relation $\partial_{\tva}=\frac{1}{\Lambda (\va)_{,\va}}
\partial_{\va}$  we get the expression for the internal energy \begin{multline} \e =\e^{eq}(\va)+\frac{\tau_{,\va}}{2\Lambda_{,\va}}\Vert q\Vert^2
-\frac{\tau(\va)}{\lambda^{0}_{,\va}}\left[ \frac{1}{2}\left(\frac{\lambda^{0}_{,\va}}{\Lambda_{,\va}} \right)_{,\va}\Vert q\Vert^2+ \left(\frac{{\tilde
K}^{A}_{,\va}}{\Lambda_{,\va}}\right)_{,\va}q^A \right]=\\ = ^{\Lambda_{,\va}=\kappa-const}\e^{eq}(\va) +\frac{\tau_{,\va}}{2\kappa }\Vert q\Vert^2
-\frac{\tau(\va)}{\kappa\lambda^{0}_{,\va}}\left[ \frac{1}{2}\lambda^{0}_{,\va\va}\Vert q\Vert^2+ {\tilde K}^{A}_{,\va\va}q^A \right]. \end{multline} Notice that for
$\lambda^{0}=0$, balance law given by the 4th column in (12.24) vanish. The same is true for deformations of the Cattaneo equation defined in the third column when
${\tilde K}^A(\va )=0$.\par

First and second balance laws in the system (12.24) are the balance laws of the original Cattaneo system. Last one is the trivial balance law.  Third and forth
columns give the balance law \begin{multline} \partial_{t}\left[ \hat{\lambda}^0 \e -\int^{\va}\lambda^{0}_{,\va} \e^{eq}ds +\tau(\va
)\Lambda^{-1}_{\va}[\frac{1}{2}\lambda^{0}_{\va}\Vert q\Vert^2+\hat{K}^{A}_{,\va}(\va)q^A]\right] +\partial_{x^A}\left[\hat{\lambda}^{0}(\va)q^A +\hat{K}^{A}(\va)
\right]=\\ = -\Lambda^{-1}_{,\va }(\hat{\lambda}^{0}_{\va}\Vert q\Vert^2 +\hat{K}^{A}_{,\va}(\va)q^A). \end{multline}

Source/production term in this equation has the form

\begin{multline}
 -\Lambda^{-1}_{,\va }(\hat{\lambda}^{0}_{\va}\Vert q\Vert^2 +\hat{K}^{A}_{,\va}(\va)q^A)=-\Lambda^{-1}_{,\va }\hat{\lambda}^{0}_{\va}(\Vert q\Vert^2
 +\frac{\hat{K}^{A}_{,\va}(\va)}{\hat{\lambda}^{0}_{\va}}q^A)=\\=
 -\Lambda^{-1}_{,\va }\hat{\lambda}^{0}_{\va}\left[\sum_{A}( q^A +\frac{\hat{K}^{A}_{,\va}(\va)}{2\hat{\lambda}^{0}_{\va}})^2 -\sum_{A} \left(
 \frac{\hat{K}^{A}_{,\va}(\va)}{2\hat{\lambda}^{0}_{\va}}\right)^2\right]
\end{multline} By physical reasons, $\Lambda_{,\va}>0.$ As (3.18) shows, $\lambda_{,\va}$ may have any sign.  We assume that this sign does not depend on $\va$.

For a fixed $\va$ expression (4.5) for the production in the balance law (4.4) may have constant sign \textbf{for all values of} $q^A$ \emph{if and only if}
$\hat{K}^{A}_{,\va}(\va)=0, A=1,2,3$. Therefore this is possible only if the internal energy (4.3) has the form \beq \e=\e^{eq}(\va)+\left[
\frac{\tau_{,\va}}{2\Lambda_{,\va}} -\frac{\tau(\va)}{2\hat{\lambda}^{0}_{,\va}} \left(\frac{\hat{\lambda}^{0}_{,\va}}{\Lambda_{,\va}} \right)_{,\va} \right] \Vert
q\Vert^2=^{\tau-const, \Lambda_{,\va}-const}\ \e^{eq}(\va) -\frac{\tau(\va)}{2k\hat{\lambda}^{0}_{,\va}} \hat{\lambda}^{0}_{,\va\va} \Vert q\Vert^2 \eeq with some
function ${\hat \lambda}^{0}(\va)$.  This being so, Cattaneo system has the supplementary balance law \begin{multline} \partial_{t}\left[ \hat{\lambda}^0 \e
-\int^{\va}\hat{\lambda}^{0}_{,\va} \e^{eq}ds +\frac{1}{2}\tau(\va )\Lambda^{-1}_{\va}\hat{\lambda}^{0}_{,\va}\Vert q\Vert^2\right]
+\partial_{x^A}\left[\hat{\lambda}^{0}(,\va)q^A \right]= -\Lambda^{-1}_{,\va }\hat{\lambda}^{0}_{,\va}\Vert q\Vert^2 \end{multline}
 \textbf{with the production term that may have constant sign - nonnegative, provided (we use the fact that $\hat{\lambda}^{0}_{,\va}={\lambda}^{0}_{,\va}$)}
 \beq
 \Lambda^{-1}_{,\va } \lambda^{0}_{,\va}\leqq 0.
 \eeq
\textbf{This inequality (which is equivalent, if $\Lambda_{,\va }\geqq 0$, to the inequality $\lambda^{0}_{,\va}\leqq 0$ ) is the II law of thermodynamics for
Cattaneo heat propagation model.}\par

If we take $\textbf{q}=0$ in the entropy balance (4.7) we have to get the value of entropy at the equilibrium $s^{el}$: \beq
s^{eq}=\hat{\lambda}^{0}\epsilon^{eq}-\int^{\va}\lambda^{0}_{,\va} \e^{eq}ds=\int^{\va}{\hat \lambda}^{0}\epsilon^{eq}_{,\va}d\va. \eeq From this it follows that
\emph{at a homogeneous state} $ds^{eq}={\hat \lambda}^{0}d\epsilon^{eq}$.  Comparing this with the Gibbs relation $d\epsilon^{eq}=\va ds^{eq}$ we conclude that \beq
\hat{\lambda}^{0}=\frac{1}{\va}. \eeq Using (3.13) we also conclude that \beq \lambda^A =-\frac{q^A}{\va^2}, A=1,2,3. \eeq It follows from this that the condition
(4.8) (II law) takes the form well known from thermodynamics (see \cite{JCL,Mu1, MR}: \beq \Lambda_{,\va}\geqq 0. \eeq

Substituting (4.10) into (4.6) and calculating $ -\frac{\tau(\va)}{2\hat{\lambda}^{0}_{,\va}}\left( \frac{\hat{\lambda}^{0}_{,\va}}{\Lambda_{,\va}}
\right)_{,\va}=\frac{\tau(\va)\va^2}{2}\left( \frac{-1}{\va^2 \Lambda_{,\va}} \right)_{,\va}=-\frac{\tau(\va)\va^2}{2}\frac{-(2\va\Lambda_{,\va} +\va^2
\Lambda_{,\va\va })}{\va^4 \Lambda^{2}_{,\va}}=\frac{\tau(\va)}{\va \Lambda_{,\va}}+\frac{\tau(\va)\Lambda_{,\va\va}}{2 (\Lambda_{,\va})^2}$ we get the expression
for internal energy in the form \beq \e=\e^{eq}(\va)+\left[ \frac{\tau_{,\va}}{2\Lambda_{,\va}}+\frac{\tau}{\va \Lambda_{\va}}+\frac{\tau \Lambda_{,\va
\va}}{2(\Lambda_{,\va})^2 } \right] \Vert q\Vert^2=^{\tau-const, \Lambda_{,\va}-const}\e^{eq}(\va)+\frac{\tau}{\va \Lambda_{,\va}} \Vert q\Vert^2. \eeq

For the entropy density we have \begin{multline} s=s^{eq}+\hat{\lambda}^{0}(\epsilon -\epsilon^{eq})+\frac{1}{2}\tau(\va
)\Lambda^{-1}_{\va}\hat{\lambda}^{0}_{,\va}\Vert q\Vert^2=\\ =s^{eq}+\frac{1}{\va}\left[ \frac{\tau_{,\va}}{2\Lambda_{,\va}}+\frac{\tau}{\va
\Lambda_{\va}}+\frac{\tau \Lambda_{,\va \va}}{2(\Lambda_{,\va})^2 } \right] \Vert q\Vert^2-\frac{\tau(\va)}{2\va^2 \Lambda_{\va}} \Vert
q\Vert^2=\\=s^{eq}+\frac{1}{\va}\left[ \frac{\tau_{,\va}}{2\Lambda_{,\va}}+\frac{\tau}{2\va \Lambda_{\va}}+\frac{\tau \Lambda_{,\va \va}}{2(\Lambda_{,\va})^2 }
\right] \Vert q\Vert^2=s^{eq}+\frac{\tau}{2\va\Lambda_{,\va}}\left[ \frac{\tau_{,\va}}{\tau}+\frac{1}{\va }+\frac{ \Lambda_{,\va \va}}{\Lambda_{,\va} } \right] \Vert
q\Vert^2=\\=^{\tau-const, \Lambda_{,\va}-const}s^{eq}+\frac{\tau}{2\va^2\Lambda_{,\va}}\Vert q\Vert^2. \end{multline}

Correspondingly, the entropy balance law takes the form

\beq \partial_{t}\left( s^{eq}+\frac{\tau}{2\va\Lambda_{,\va}}\left[ \frac{\tau_{,\va}}{\tau}+\frac{1}{\va }+\frac{ \Lambda_{,\va \va}}{\Lambda_{,\va} } \right]
\Vert q\Vert^2 \right) +\partial_{x^A}(\frac{q^A}{\va})=\frac{1}{\Lambda_{,\va}} \Vert \frac{\mathbf{q}}{\va}\Vert^2. \eeq \begin{remark} If in the absence of the
heat flow ($\textbf{q}=0$) the "equilibrium state" is not homogeneous, more general constitutive relations with $\lambda^{0}$ different from (5.10) and more general
form of energy and entropy entropy balances  satisfying to the II law of Thermodynamics, are possible. \end{remark} \vskip1cm We collect obtained results in the
following \begin{theorem} \begin{enumerate} \item For the Cattaneo heat propagation balance system (2.1) compatible with the entropy principle and having a
nontrivial supplementary balance law that is not a
    linear combination of the original balance laws with constant coefficients, the internal energy has the form (4.3).
 If (4.3) holds, all supplementary balance laws for Cattaneo balance system (including original equations and the trivial ones) are listed in (4.2).  New
 supplementary
 balance laws depend on the 4 functions of temperature - $\hat{\lambda}^{0}(\va),\tilde{K}^{A}(\va), A=1,2,3$.  Corresponding main fields $\lambda^\mu,\mu=0,1,2,3$
 have the form (4.10-11).\par
\item Additional balance law (4.4) given by the sum of third and forth columns in (4.2) \emph{has the nonnegative production term} if and only if the internal
    energy
    $\epsilon$ has the form (4.12) and, in addition, the condition (4.11) holds. Cattaneo systems satisfying to these conditions depend on one arbitrary function of
    time $\epsilon^{eq}(\va )$.
\item Supplementary balance law having nonnegative production term (entropy) is unique modulo linear combination of original balance laws and the trivial balance
    laws.
\end{enumerate} \end{theorem}

\section{Conclusion.}

Description of the supplementary balance laws for Cattaneo heat propagation system given in this paper can probably be carried over for other systems of balance
equations for the couples of fields: scalar + vector field. \par One observes a kind of duality between the tensorial structure of dynamical fields (here $\theta,
\textbf{q}$) and the list of free functions of temperature $\lambda^0 (\va), {\tilde K}^{A}(\va ), A=1,2,3$ entering the description of SBL. It would be interesting
to follow up if similar duality exists for the balance systems of more complex tensorial structure (description of SBL of lowest order (=1) was done for the
Navier-Stokes-Fourier fluid, see \cite{Pr2}).\par II law of thermodynamics - positivity of production determines the entropy balance uniquely, modulo addition of
trivial balance laws and the linear combination of the original balance laws.  It would be interesting to formulate mathematical conditions removing even this
trivial non-unicity and delivering some "optimal" form of the entropy balance.


\vskip0.4cm


\begin{thebibliography}{99}

\bibitem{CA} H. Callen, \emph{Thermodynamics}, Whiley, 2nd ed. 1985.

\bibitem{GP}  P.Glensdorf, I.Prigogine, \emph{Thermodynamical Theory of Structure, Stability and fluctuations}, Wiley, Brussels, 1971.

\bibitem{JCL} D.Jou, J.Casas-Vasquez, G.Lebon, \emph{Extended Irreversible Thermodynamics}, 3rd ed., Springer, 2001.

\bibitem{L}  I-Shish Liu. Method of Lagrange multipliers for exploatation of the entropy principle, Arch. Rational Mech. Anal., v.46, 1972, pp.131-148.

\bibitem{MR} I. Muller, T. Ruggeri, \emph{Rational Extended Thermodynamics}, 2nd ed., Springer, 1998.

\bibitem{Mu} I.Muller, \emph{Thermodynamics}, Pitman Adv. Publ., Boston, 1985. co.,1985.

\bibitem{Pr} S. Preston, \emph{Geometrical Theory of Balance Systems and the Entropy Principle},
 Proceedings of GCM7, Lancaster, UK, Journal of Physics: Conference Series, vol.62, pp.102-154, 2007.

\bibitem{Pr1} S.Preston, "Variational theory of balance systems", Intern. J. of Geom. Methods of Modern Phys., v7, N5 (August) 2010.

\bibitem{Pr2} S.Preston, Supplementary balance laws for the navier-Stokes-fourier Fluid, Manuscript, unpublished.

\bibitem{R1} T. Ruggeri, \emph{Galilean Invariance and Entropy Principle For Systems of Balance Laws}, Cont. Mech.Thermodyn. 1 (1989).

\bibitem{R3} T. Ruggeri, \emph{The Entropy Principle: from Continuum Mechanics to Hyperbolic Systems of Balance Laws}, Entropy, v.10, pp.319-333, 2008.

\bibitem{PR} S. Pennisi, T. Ruggeri, \emph{A new method to exploit the entropy Principle and galilean invariance in the macroscopic approach to Extended
    Thermodynamics},
    Ricerche di Matematica, 55, 2006, pp. 319-339.

\bibitem{Ser} D. Serre,\emph{Systems of Conservation Laws I} CUP, Cambridge, 1999.

\bibitem{TN} C. Truesdell, W. Noll, \emph{The Non-Linear Field Theories of Mechanics}, 2nd ed., Springer, 1992. \end{thebibliography}
\end{document}